\documentclass[aps,prb,twocolumn,showpacs,amsmath,twocolumn,amssymb,superscriptaddress,letterpaper]{revtex4}
\usepackage{times}
\usepackage{amsfonts}
\usepackage{mathrsfs}
\usepackage{graphicx}% Include figure files
\usepackage{dcolumn}% Align table columns on decimal point
\usepackage{bm}% bold math
\usepackage{color}
\usepackage{subfigure}

\usepackage[colorlinks,bookmarks=false,citecolor=blue,linkcolor=red,urlcolor=blue]{hyperref}
\begin{document}
\title{ Topological phase in non-centrosymmetric material NaSnBi }

\author{Xia Dai}
\affiliation{Institute of Physics, Chinese Academy of Sciences, Beijing 100190, China}

\author{Congcong Le}
\affiliation{Institute of Physics, Chinese Academy of Sciences, Beijing 100190, China}

\author{Xianxin Wu}
\affiliation{Institute of Physics, Chinese Academy of Sciences, Beijing 100190, China}

\author{Shengshan Qin}
\affiliation{Institute of Physics, Chinese Academy of Sciences, Beijing 100190, China}

\author{Zhiping Lin}
\affiliation{Institute of Physics, Chinese Academy of Sciences, Beijing 100190, China}

\author{Jiangping Hu  }
\affiliation{Institute of Physics, Chinese Academy of Sciences, Beijing 100190, China}
\affiliation{Collaborative Innovation Center of Quantum Matter, Beijing, China}

\date{\today}

\begin{abstract}
We predict that a non-centrosymmetric material NaSnBi locates in a three-dimensional non-trivial topological phase under ambient pressure based on first-principle calculations. By deriving the effective model around $\Gamma$ point, we find that the topological phase transition is driven by a Rashba spin-orbital coupling through an odd number of pairs of band touch because of a small anisotropic gap caused by quintic dispersion terms. In contrast to conventional topological insulators, the spin texture of surface Dirac cone is right-handed and the surface states are strikingly different for different surface terminations.
\end{abstract}


\maketitle

Since the theoretical discovery of topological insulators (TIs)\cite{Kane2005,Fuliang2007prl,Bernevig2006}, they have attracted tremendous interests in condensed matter physics\cite{Moore2010}. The most striking property of TIs is the emergence of symmetry-protected metallic surface or edge states while the bulk bands are fully gapped. HgTe/CdTe quantum well was theoretically predicted to be the first TI and then it was confirmed by transport experiments\cite{Bernevig2006,Markus2007}. Soon afterwards, another two dimensional (2D) quantum well InAs/GaSb was also theoretically discovered\cite{Liu2008,Knez2011,Spanton2014}. Other 2D TIs that have been discovered by first principle calculations include bismuth bilayers\cite{Murakami2006}, monolayer Bi$_4$Br$_4$\cite{Zhou2014}, single-layer ZrTe$_5$ and HfTe$_5$\cite{Weng2014}. Three dimensional (3D) TIs were first predicted in Bi$_{1-x}$Sb$_x$\cite{Teo2008}, followed by Bi$_2$Se$_3$, Bi$_2$Te$_3$ and Sb$_2$Te$_3$\cite{Haijun2009}, which have large band gaps. The non-trivial surface state, Dirac cone, was observed experimentally in the angle-resolved photoemission spectroscopy (APRES) experiments in Bi$_2$Se$_3$\cite{Hsieh2008,Hsieh2009,Wray2010}. The key feature of topologically protected surface states is the spin texture, which gives rise to a non-trivial Berry phase and forbids back scattering, and it has been observed in experiment\cite{Hsieh2009}. In the above systems, most of them are centrosymmetric and the spin-orbit coupling (SOC) plays an essential role. The topological phase transition occurs through a band inversion at time reversal invariant momenta\cite{Fuliang2007,Binhai2010,Feng2011,Xxwu2016}.

In the system with inversion symmetry, even with SOC the bands are still spin-degenerate as long as time reversal symmetry is not broken. Breaking inversion symmetry can lift the spin degeneracy through Rashba spin splitting and is also important for many exotic phenomena, for example multiferroics and Weyl semimetals\cite{Weng2015,BQLv2015,Huang2015,SYXu2015,BQLv2015b}. It is essential for the appearance of ferroelectric polarization and the separation of Weyl points with opposite chiral charge. Moreover, without definite parity, in the superconducting state, the Cooper pair in principle can be the mix of spin singlet and spin triplet states. In this case, with external magnetic field, topological superconductivity may be realized, which can host Majorana fermions\cite{Jason2012,Fuliang2008,Linder2010}. Realizing topologically non-trivial phase  in a noncentrosymmetric system can help us to integrate topology and the above features in a single material. For example, the layered material BiTeI under relative large external pressure is predicted to possess large bulk Rashba SOC and to be a topological insulator\cite{Bahramy2012}.

\begin{figure}
\centerline{\includegraphics[width=0.48\textwidth]{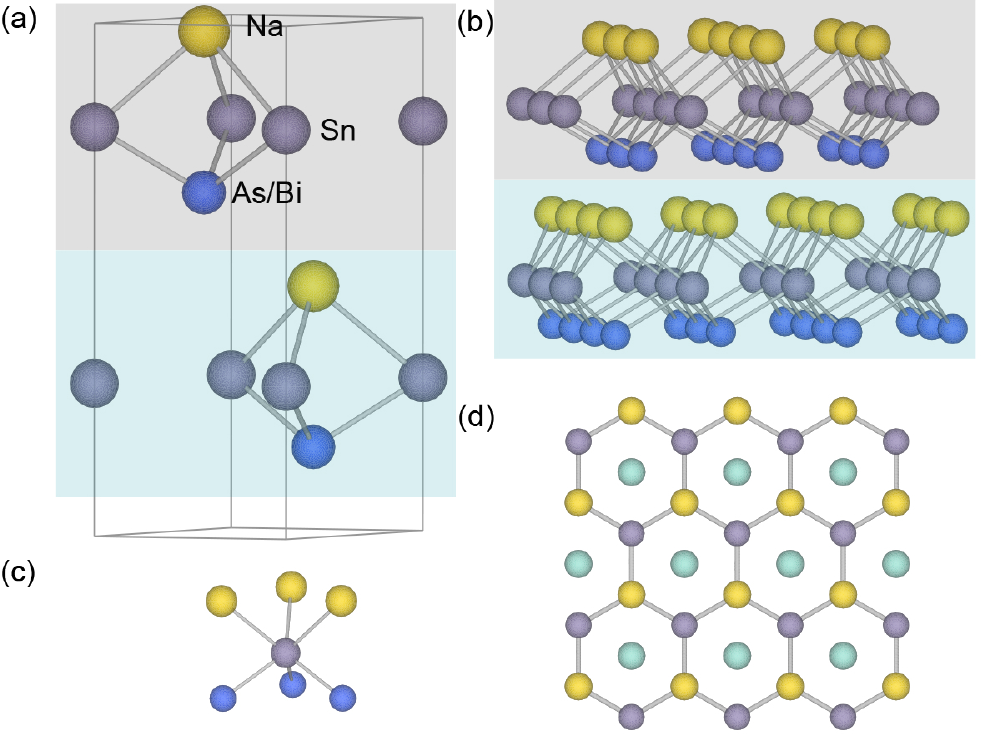}} \caption{(color online) \textbf{Crystal structure of NaSnX (X=As, Sb, Bi).} (a) Side view of a unitcell of NaSnX. The shades of grey and blue denote A and B sublattice, respectively. (b) 3D view of the structure. (c) A Na-Sn-X trigonal prismatic geometry. (d) Top view of the structure, in which the yellow and green Na atoms are located at upper A and downer B sublattice, respectively.
 \label{structure} }
\end{figure}

In this paper, we perform first-principle calculations to investigate the electronic structures of non-centrosymmetric materials NaSnX (X=As, Sb, Bi) and predict that NaSnBi is a metal in a 3D $Z_2$ topologically non-trivial phase under ambient pressure. Through the calculations of Wannier charge centers (WCCs), we confirm that the topological invariant is $Z_2=1;(000)$.  By deriving the effective \textbf{k $\cdot$ p} model near $\Gamma$ point, we find that the quintic dispersion terms can cause a small anisotropic gap and are responsible for the topological phase transition that occurs through an odd number of pairs of band touch. The topologically protected surface states strongly depend on termination directions. Furthermore, the spin texture of the surface state is right-handed in sharp contrast to conventional TIs. The materials can be an ideal system to realize topological superconductivity by integrating both topology and superconductivity.

Our calculations are performed using density functional theory (DFT) employing the projector augmented wave (PAW) method encoded in the Vienna \emph{ab initio} simulation package (VASP) \cite{Kresse1993,Kresse1996,Kresse1996b}. Generalized-gradient approximation (GGA)\cite{Perdew1996} for the exchange correlation functional is used. The cutoff energy is set to be 600 eV for expanding the wave functions into plane-wave basis for structural optimization and 450 eV for static and band calculations. The Brillouin zone is sampled in the $\textbf{k}$ space within Monkhorst-Pcak scheme\cite{Monkhorst1976} and the number of these $\textit{k}$ points is ${12 }\times{ 12 }\times{ 4}$. We relax the lattice constants and internal atomic positions for NaSnBi and NaSnSb with GGA, where forces are minimized to less than 0.01 eV/\AA. Real-space force constants of supercells are calculated in the density-functional perturbation theory (DFPT)\cite{Baroni1987}, and phonon frequencies are calculated from the force constants using the PHONOPY code\cite{Togo2008,Togo2015}.

NaSnX (X=As, Sb, Bi) belongs to the hexagonal space group of \emph{P6$_3$mc} and has a non-centrosymmetric layered structure along the $z$ direction, as illustrated in Fig.\ref{structure}. In a unitcell, there are two alternately stacked Na-Sn-X trilayers, where Sn and X atoms form strong covalent bonds and Sn and Na atoms form ionic bonds, which make the trilayer possess an intrinsic polar axis along the $z$ direction. The trigonal prismatic coordination in trilayer is similar to that in the 2H-MoS$_2$ although the top and bottom atoms are Na and X atoms instead of both S atoms. The two trilayers are related to each other by a screw rotation $\bar{C}_{6z}$ ($\bar{C}_{6z}= \{C_{6z}|\textbf{t}\}$), a sixfold rotation around \emph{c} axis followed by a fractional lattice transition \textbf{t} ($\textbf{t}=\textbf{c}/2$), where \textbf{c} is a lattice basis vector, with the position of Sn atoms being chosen as the spatial origin. In contrast to MoS$_2$, the atoms above and below Sn layers are different, making the inversion-symmetry break in NaSnX crystal structure.

Firstly we investigate the stability of NaSnX. As NaSnAs has been experimentally synthesized\cite{NaSnAs}, we substitute As with Sb and Bi in the same group. The crystal structures of NaSnSb and NaSnBi are fully relaxed and the obtained lattice parameters are listed in Tab. \ref{tab_parameter}. Compared with NaSnAs, the lattice constant $c$ changes a little but $a$ increases by 9\% and 12.5\% for NaSnSb and NaSnBi, respectively. It is well known that the binding energy is very useful for estimating the stability of new crystal structures. Here, the binding energy per atom, $E_b$, is defined as $E_b=(2E_{Na}+2E_{Sn}+2E_{X}-E_{total})/6$~(X=As, Sb, Bi), in which $E_{Na}$, $E_{Sn}$, $E_{As}$, $E_{Sb}$ and $E_{Bi}$ are the respective energies per atom of element Na, Sn, As, Sb and Bi in the states at standard ambient temperature and pressure. $E_{total}$ is the total energy of a unitcell. The obtained binding energy is 441 meV and 382 meV per atom for NaSnSb and NaSnBi, respectively, which are smaller compared with that of NaSnAs, 524 meV per atom, indicating that NaSnSb and NaSnBi are favorable in energy in experiment and may be synthesized using similar methods with NaSnAs. To further address the stability of the crystal structure, we calculate the phonon spectrums. No imaginary frequencies are observed throughout the whole Brillouin zone for both NaSnSb and NaSnBi (see supplementary materials), confirming their dynamically structural stability. Through the calculations of binding energy and phonon spectrums, we may conclude that NaSnSb and NaSnBi are stable and may be synthesized in future experiment.

\begin{table}
\caption{Experimental structural parameters of NaSnAs and optimized structural parameters of NaSnSb and NaSnBi, in \emph{p6$_3$mc} space group.}
\label{tab_parameter}
\begin{ruledtabular}
\begin{tabular}{ccccc}
     R     &                   &  NaSnAs  &  NaSnSb  &  NaSnBi  \\
\colrule
 $a$(\AA)  &                   &   4.001  &   4.361  &   4.500  \\
 $b$(\AA)  &                   &   4.001  &   4.361  &   4.500  \\
 $c$(\AA)  &                   &  11.729  &  11.836  &  11.695  \\
     Na    & 2$b$(1/3 2/3 $z$) &          &                     \\
    $z$    &                   &  0.9955  &  0.9859  &  0.9805  \\
 As/Bi/Sb  & 2$b$(1/3 2/3 $z$) &          &                     \\
    $z$    &                   &  0.6660  &  0.6666  &  0.6666  \\
     Sn    &   2$a$(0 0 $z$)   &          &                     \\
    $z$    &                   &  0.2836  &  0.2926  & 0.2980  \\
\end{tabular}
\end{ruledtabular}
\end{table}

\begin{figure}
\centerline{\includegraphics[width=0.48\textwidth]{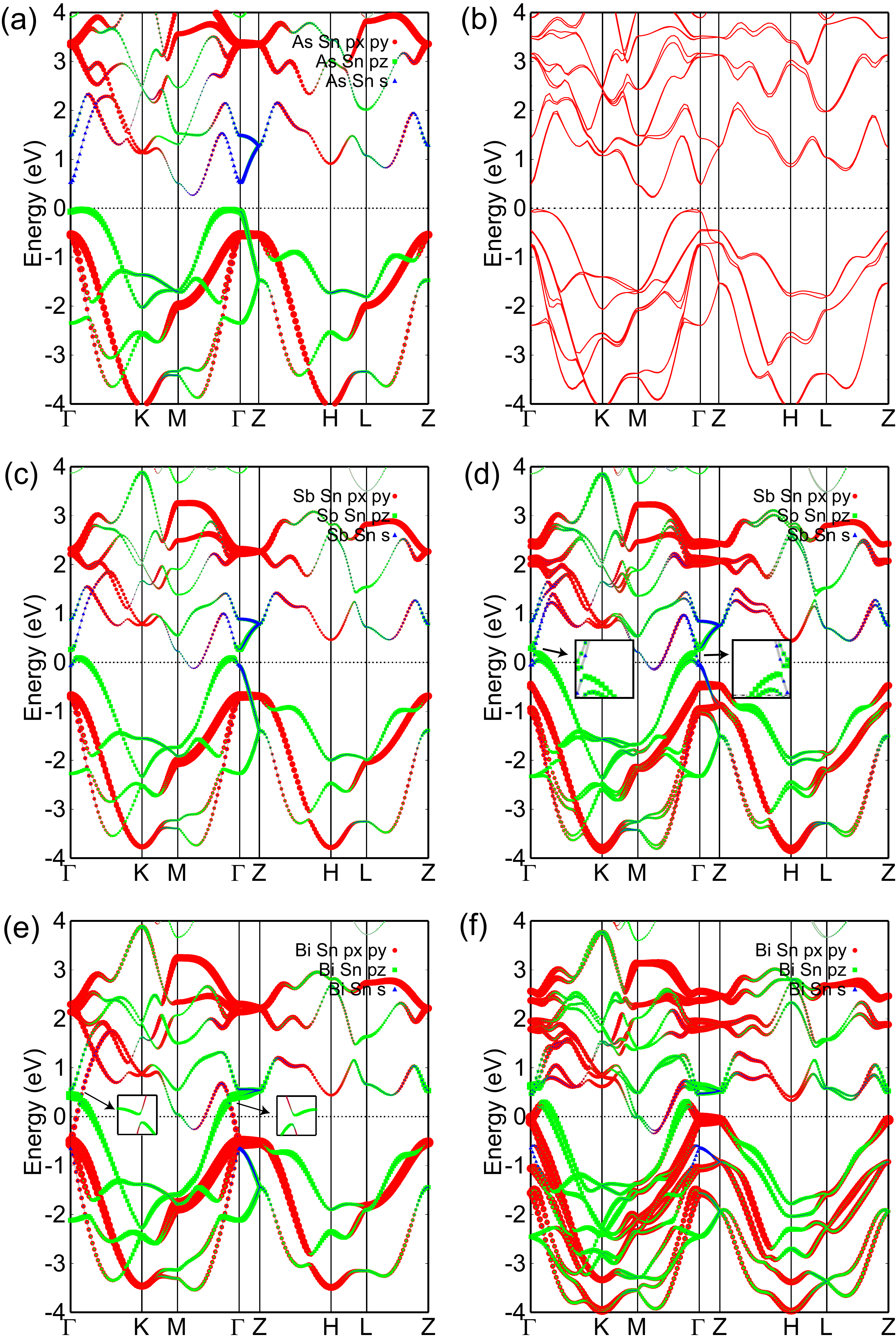}} \caption{(color online) Calculated band structures of NaSnAs ((a),(b)), NaSnSb ((c),(d)), NaSnBi ((e),(f)) without ((a),(c),(e)) and with SOC ((b),(d),(f)).
 \label{band_NaSnX} }
\end{figure}

In order to investigate the topological properties of this family of materials, we analyze the electronic structures of NaSnX. NaSnAs is a narrow-gap semiconductor, with an indirect gap of 0.3 eV, as shown in Fig.\ref{band_NaSnX} (a). The conduction band minimum locates at $\Gamma$--$M$ line near $M$ point and the valence band maximum is at $\Gamma$ point, near which the band is relatively flat. The direct bandgap at $\Gamma$ point is about 0.4 eV. The conduction bands are mainly attributed to the $s$ orbitals of Sn and As atoms, while the valence bands are contributed by the $p$ orbitals of As atoms. Furthermore, the topmost valence band is attributed to As $p_z$ orbital and the valence bands below are attributed to $p_x$ and $p_y$ orbitals due to the crystal field. From the orbital characters of conduction bands and valence bands, we know that the first order of SOC vanishes. Moreover, due to the small onsite SOC strength of As, we can expect that introducing SOC will have little effects on the gap of NaSnAs, as shown in Fig.\ref{band_NaSnX} (b). Because of the absence of inversion symmetry, it is not possible to assign a parity to each band at the time reversal invariant momenta. According to the calculations of WCCs\cite{Alexey2011}, we find that NaSnAs is topologically trivial.

Then we consider the band structure of NaSnSb, shown in Fig.\ref{band_NaSnX} (c). It is metallic but has a local gap. Compared with NaSnAs, the most noticeable feature is that there is a band inversion at $\Gamma$ point between the $s$-character state and $p_z$-character state of Sn and Sb atoms. In centrosymmetryic material systems, the band inversion between two states with opposite parities will change $Z_2$ topological indices. However, inversion symmetry is broken in our case. Our further calculations of topological invariants show that NaSnSb is also topologically trivial. Besides the band inversion, there are large Rashba splittings for both conduction and valence bands near $\Gamma$ point due to the large SOC of Sb and the gap between them is very small, as shown in Fig.\ref{band_NaSnX} (d). Further enhancing the strength of SOC, we find that the gap can be closed along $\Gamma$--$K$ and $\Gamma$--$M$ lines. If the band touch just happens in either $\Gamma$--$M$ or $\Gamma$--$K$ direction, a topologically nontrivial phase can emerge.

Now we consider the case of NaSnBi, where the onsite SOC strength is further enhanced. The band structures are given in Fig.\ref{band_NaSnX} (e) and (f). The band inversion is further enhanced and the $s$-character state sinks below the $p_{x,y}$ state. The splitting in $p_x,~p_y$ bands and Rasha splitting near $\Gamma$ point are much larger compared with NaSnSb due to the larger SOC of Bi and the involvement of $p_{x,y}$ orbitals. According to our calculations (see the supplementary materials), the $Z_2$ topological indices of NaSnBi are 1;(000), which indicates that it is in a strong topologically nontrivial phase.

\begin{figure}
\centerline{\includegraphics[width=0.48\textwidth]{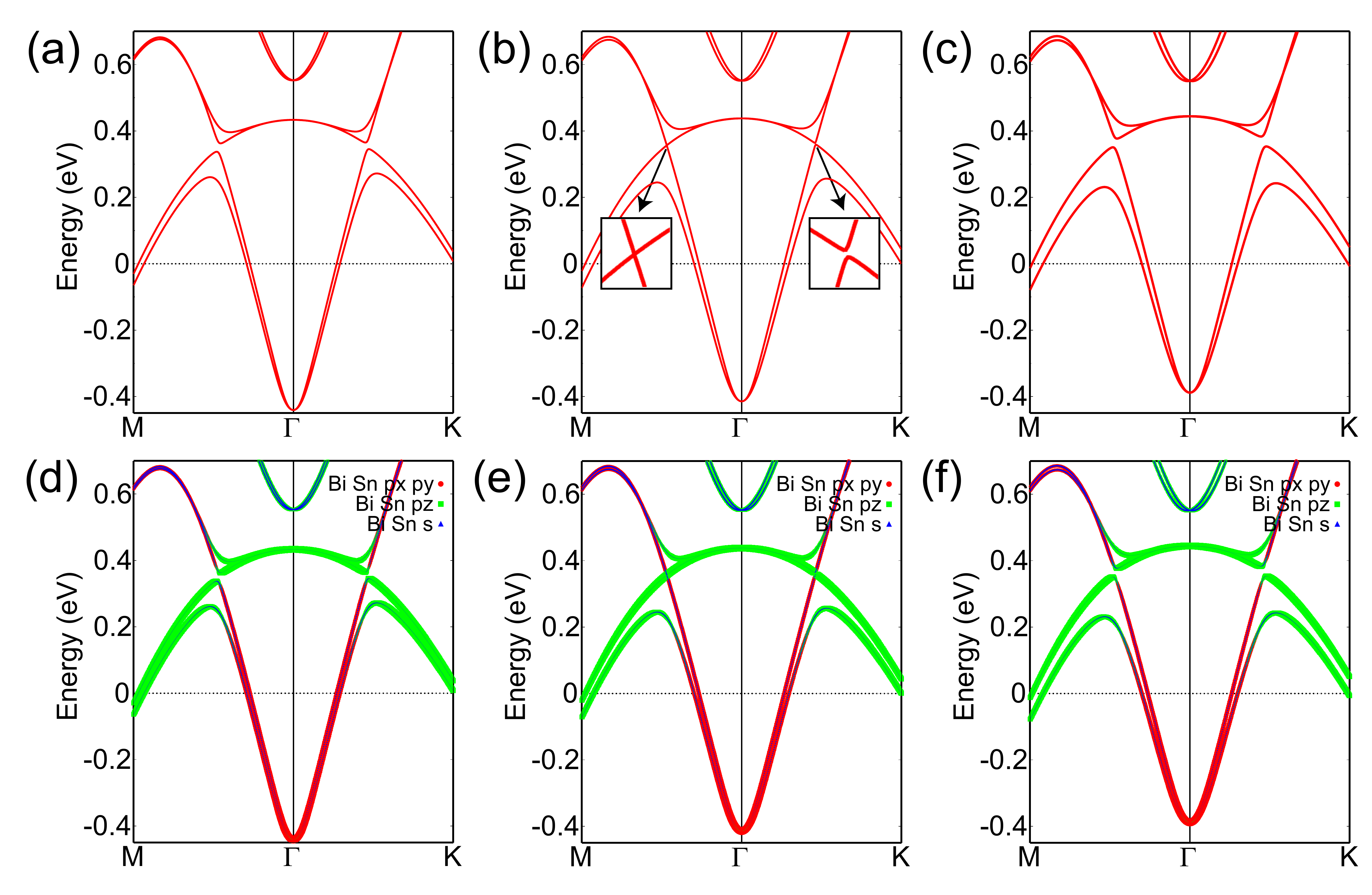}} \caption{(color online) \textbf{Effect of adjusting the strength of SOC on the bulk electronic states for NaSnBi.} Band structures of NaSnBi with (a) 10\%, (b) 15\% and (c) 20\% of the real atomic SOC. The orbital characteristics of (a), (b) and (c) are shown in (d), (e) and (f), respectively.
 \label{band_soc10_15_20} }
\end{figure}

How does the topological phase transition happen? To elucidate the mechanism, we perform calculations by adjusting the strength of SOC ($\lambda_{Bi}$) of Bi from 0 to the real atomic value. Let us focus on the gap near $\Gamma$ point, the bands exhibit Rashba splitting when SOC is turned on, as shown in Fig.\ref{band_soc10_15_20} (a). The gap $\Delta_K (\Delta_M)$ along $\Gamma$--$K$ ($\Gamma$--$M$) direction decreases with the increasing of SOC. $\Delta_K$ and $\Delta_M$ are close but $\Delta_K$ is a bit larger. When the SOC strength is $\lambda_c$, 15\% of the real atomic value, there is a band touch along $\Gamma$--$M$ direction but no in the $\Gamma$--$K$ direction, as shown in Fig.\ref{band_soc10_15_20} (b). Further increasing the SOC strength will reopen a gap but the gap along $\Gamma$--$K$ is never closed in the whole process, as shown in Fig.\ref{band_soc10_15_20} (c). As the band touch is mediated through an odd number of pairs (just along $\Gamma$--$M$ direction), this phase transition should in principle change the $Z_2$ indices. The calculations for $Z_2$ invariant confirm our conjecture. The system stays in a topologically trivial phase for $\lambda_{Bi}<\lambda_c$ but becomes topologically non-trivial ($Z_2$=1;(000)) for $\lambda_{Bi}>\lambda_c$. Moreover, we have perform calculations for NaSnSb with unreal large SOC (180\% of the atomic value) and found similar phase transition. The results indicate that the systems can undergo a topological phase transition with the increasing of SOC strength.

Without SOC there has already been a band inversion and SOC will not introduce additional band touch in NaSnSb, we usually expect that it is topologically non-trivial just like conventional TIs but it is trivial. Compared with NaSnAs, the band touch is mediated through even number of pairs (both along $\Gamma$--$M$ and $\Gamma$--$K$ directions) in NaSnSb, therefore, the $Z_2$ topological phase transition is not allowed.

\begin{figure}
\centerline{\includegraphics[width=0.48\textwidth]{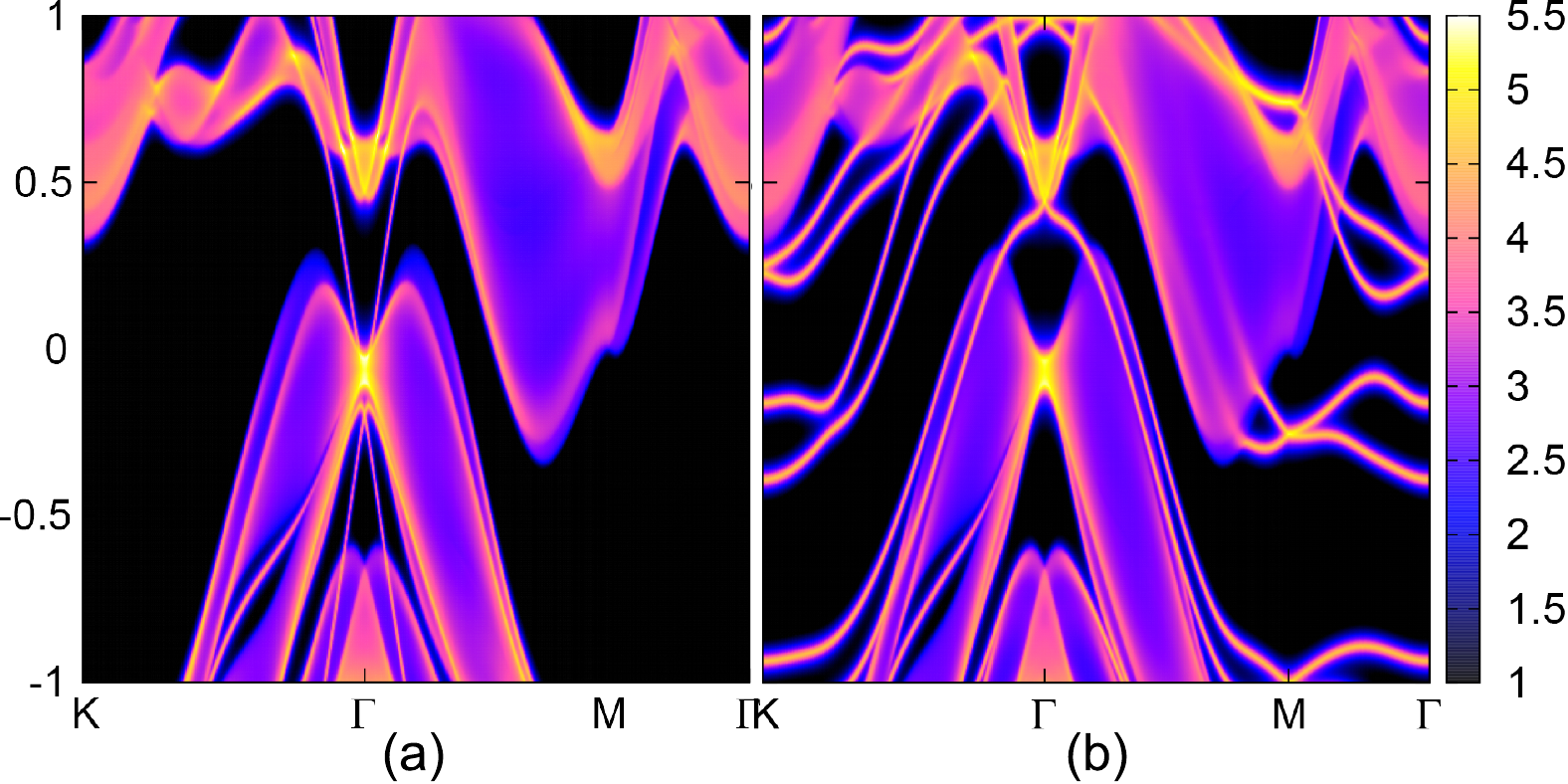}} \caption{(color online) Energy and momentum dependence of LDOS of (a) Sn-terminated and (b) Bi-terminated (001) edge for NaSnBi.
 \label{edgeband_NaSnBi} }
\end{figure}

\begin{figure}
\centerline{\includegraphics[width=0.3\textwidth]{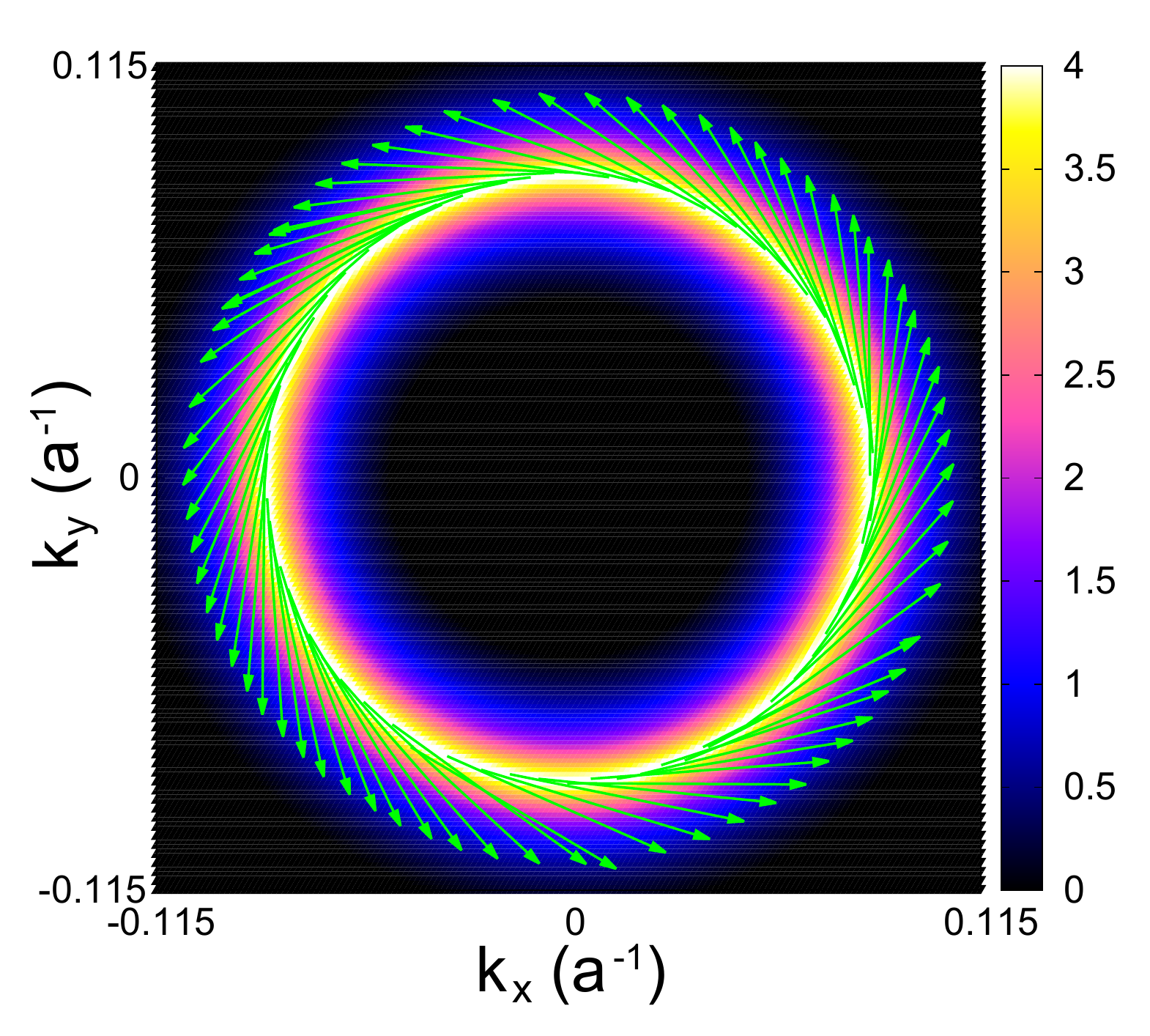}} \caption{(color online) The fermi surface (E$_f$=100meV) and spin texture for the Sn-terminated (001) surface state of NaSnBi.
 \label{spintexture} }
\end{figure}

The physical consequence of non-trivial topology is the appearance of gapless surface states. To obtain the surface states, we calculate the surface Green's function of the semi-infinite system using an iterative method\cite{Lopez1984,Lopez1985}. The edge local density of states (LDOS) of Sn-terminated and Bi-terminated (001) surface for NaSnBi are shown in Fig.\ref{edgeband_NaSnBi}. The surface states are quite different: the Dirac point is buried in bulk valence bands in the former case and in bulk conduction bands in the latter case. The Fermi surface of the Sn-terminated surface state (E$_f$=100meV) is shown in Fig.\ref{spintexture} and the spin orientation for the surface state around the Fermi surface is marked by green arrows. While moving around the Fermi surface, the spin orientation rotates simultaneously, forming a right-handed spin-orbit ring, which indicates the bulk non-trivial topology. The spin texture is opposite to that of conventional TIs. Because the surface is very close to the Fermi level, it may be detected in future ARPES experiment.

To further understand the mechanism of the topological phase transition, we construct the \textbf{k $\cdot$ p} model Hamiltonian near $\Gamma$ point. In order to employ the theory of invariants, we first consider all the symmetry operations for these systems. The spatial symmetries of NaSnX are described by the non-symmorphic space group $\bar{C}_{6v}$, which is generated by two elements: a screw rotation $\bar{C}_{6z}$ and a glideless reflection $M_y : (x,y,z)\rightarrow(x,-y,z)$. In NaSnX systems, the screw operation $\bar{C}_{6z}$ just exchanges the two sublattices in two layers for each element. Another important symmetry is the time-reversal symmetry $T$. As analyzed above, the gap closing occurs between bands derived from the $s$ and $p_z$ orbits of X and Sn atoms. It is natural to consider only these two orbits as the basis in the effective Hamiltonian. Our DFT calculations show that the two states at $\Gamma$ point both belong to $J_z=\frac{5}{2}$ states. Thus, the basis with SOC at $\Gamma$ point can be written as
\begin{eqnarray}
 |\alpha,\sigma\rangle=\frac{1}{\sqrt{2}}(|A,\alpha,\sigma\rangle-|B,\alpha,\sigma\rangle),
\end{eqnarray}
where $\alpha$ and $\sigma$ denote orbital and spin, respectively ($\alpha=s, p_z$ and $\sigma=\uparrow, \downarrow$), and $A,~B$ represent the sublattices for X and Sn atoms. According to the theory of invariant, the effective Hamiltonian should preserve both spacial symmetry and time reversal symmetry. The obtained 4$\times$4 effective Hamiltonian around $\Gamma$ point up to $O(k^5)$ yields
\begin{eqnarray}
H_{eff}(\mathbf{k})& = &\epsilon(\mathbf{k})+ \mathcal{M}(\mathbf{k})\Gamma_{45}+\mathcal{Z}(\mathbf{k})\Gamma_4 \nonumber \\
                   &&  +\nu_1(1+\alpha_1(k_x^2+k_y^2))(\Gamma_1 k_y-\Gamma_2 k_x)  \nonumber \\
                   &&  +\nu_2(1+\alpha_2(k_x^2+k_y^2))(\Gamma_{23} k_y-\Gamma_{31} k_x)  \nonumber \\
                   &&  +\nu_3(1+\alpha_3(k_x^2+k_y^2))(\Gamma_{14} k_y-\Gamma_{24} k_x)  \nonumber \\
                   &&  +H_5(\mathbf{k}) \\
H_5(\mathbf{k}) &=& \left(\begin{array}{cccc}
0 & \mathcal{L}(\mathbf{k}) & A_1(\mathbf{k}) & A_2(\mathbf{k}) \\
\mathcal{L}(\mathbf{k}) & 0 & A_2(\mathbf{k}) & A_3(\mathbf{k}) \\
A_1(\mathbf{k})^* & A_2(\mathbf{k})^* & 0 & \mathcal{L}(\mathbf{k}) \\
A_2(\mathbf{k})^*& A_3(\mathbf{k})^* & \mathcal{L}(\mathbf{k}) & 0  \\
\end{array}\right),
\end{eqnarray}
where $k_{\pm}=k_x\pm ik_y$, $\epsilon(\mathbf{k})=\epsilon_0+\epsilon_1k_+k_-+\epsilon_2k_+^2k_-^2+\epsilon_3k_z^2+\epsilon_4k_z^4$, $\mathcal{M}(\mathbf{k})=M_0+M_1k_+k_-+M_2k_+^2k_-^2+M_3k_z^2+M_4k_z^4$, $\mathcal{L}(\mathbf{k})=L_0+L_1k_+k_-+L_2k_+^2k_-^2$, $\mathcal{Z}(\mathbf{k})=Z_1k_z+Z_2k_z^3+Z_3k_z^5$ and $A_i(\mathbf{k})=i(A_{i1}k_+^5+A_{i2}k_+^2k_-^3) (i=1, 2, 3)$. $M_0M_1<0$ and $M_0M_1+L_0L_1<0$ correspond to the regime of band inversion, while the other cases correspond to the normal regime. Here the band inversion does not mean non-trivial topology. The appearance of $A_1(\mathbf{k})$ and $A_3(\mathbf{k})$ is due to the absence of inversion symmetry in these systems. As the gap closing occurs in $k_z=0$ plane, we only consider the model for $k_z=0$ in the following. In principle, all of the coefficients can be determined by fitting the energy spectrum of the Hamiltonian with that of DFT calculations.

To clearly see the difference between gaps along $\Gamma$--$M$ and $\Gamma$--$K$, we simplify the model and focus on either conduction or valence bands. It is reasonable because the effective models for conduction and valence bands are similar. A 2$\times$2 Hamiltonian based on $s$ ($p_z$) orbital can be constructed to describe the conduction (valence) bands. Up to quintic order terms of $\mathbf{k}$, it turns out to be
\begin{equation}
H_{s}(\mathbf{k})  = \left(\begin{array}{cc}
C(\mathbf{k}) & B(\mathbf{k}) \\
B(\mathbf{k})^* & C(\mathbf{k}) \\
\end{array}\right),
\end{equation}
where $C(\mathbf{k})=C_0+C_1k_+k_-+C_2k_+^2k_-^2$ and $B(\mathbf{k})=i(B_1k_+^5+B_2k_+^2k_-^3+B_3k_+k_-^2+B_4k_-)$. For convenience of the analysis, we rewrite the Hamiltonian in the following form
\begin{eqnarray}
H_s(\mathbf{k})  = C(\mathbf{k})+[B_4+B_3(k_x^2+k_y^2)](k_y\sigma_x-k_x\sigma_y)\nonumber\\
-(B_1+B_2)k_x(k_x^2-3k_y^2)[2k_xk_y\sigma_x+(k_x^2-k_y^2)\sigma_y]\nonumber\\
-(B_1-B_2)(3k_x^2-k_y^2)k_y[(k_x^2-k_y^2)\sigma_x-2k_xk_y\sigma_y].
\end{eqnarray}
The second term in $H_s(\mathbf{k})$ is obversely the Rashba term, contributing to an isotropic in-plane spin splitting. The third and forth term act as warping terms, hexagonally distorting the energy bands. By solving the eigenvalues of the 2$\times$2 Hamiltonian, we obtain the eigenenergies along the directions $\Gamma$--$K$ and $\Gamma$--$M$
\begin{eqnarray}
E_{\Gamma-K}(\mathbf{k}) = C(k_y)\pm [B_4k_y+B_3k_y^3+(B_2-B_1)k_y^5]\\
E_{\Gamma-M}(\mathbf{k}) = C(k_x)\pm [B_4k_x+B_3k_x^3+(B_2+B_1)k_x^5],
\end{eqnarray}
where $C(k)=C_0+C_1k^2+C_2k^4$. The two eigenenergies are the same up to quartic terms of $\mathbf{k}$ and differ at quintic order terms only. The terms $B_1$--$B_4$ appear due to the conservation of angular momentum and are in principle all nonzero. As long as $B_1\neq0$, the two eigenenergies are close but different due to the quintic terms. The case of the valence bands is similar. Consequently, the energy gaps along $\Gamma$--$K$ and $\Gamma$--$M$ are different at the whole range of $\lambda_{soc}$: $\Delta_K>\Delta_M$ for $B^c_1<0$ and $\Delta_K<\Delta_M$ for $B^c_1>0$ if $B_1$ for conduction and valence bands are opposite ($B^c_1B^v_1<0$). Thus, the band touch would only occur along one of the directions, $\Gamma$--$M$ for $B^c_1<0$ and $\Gamma$--$K$ for $B^c_1>0$, which guarantees the number of the pairs of the band touch $k$-points is odd and can induce a topological phase transition. The effect of increasing $\lambda_{soc}$ is to enhance the coefficients of Rashba and warping terms. According to our fitting, $B_1$ is about 400 eV$\cdot$\AA$^5$ and the gap anisotropy is about 0.5 meV, which is very small but consistent with DFT calculations. As we adjust the strength of Rashba SOC in the effective model, we find similar behaviors with DFT calculations (see the supplementary materials).

These materials may exhibit many interesting properties. For example, it has been argued that $Z_2$-even and $Z_2$-odd phase of a noncentrosymmetric insulator must always be separated by a Weyl semimetal phase\cite{Liu2014}. Therefore, we may look for a Weyl semimetal phase in the above systems with substitution or external pressure. The insulating materials may exhibit ferroelectricity. NaSnSb and NaSnBi, which are metallic, may be superconducting at low temperature. Thus, the materials can achieve a natural integration of superconductivity and non-trivial topology to realize topological superconductivity and Majorana Fermion.

In conclusion, we predict that NaSnBi is in a 3D $Z_2$ topologically non-trivial phase. By constructing the effective model around $\Gamma$ point, we find that the topological phase transition is driven by the strong Rashba SOC through an odd number of pairs of band touch because of a small anisotropic gap caused by quintic dispersion terms. The surface state possesses a "right-handed" spin texture in sharp contrast to conventional TIs.

{\it Acknowledgments:}
The work is supported by the Ministry of Science and Technology of China 973 program( No. 2015CB921300), National Science Foundation of China (Grant No. NSFC-1190020, 11334012), and the Strategic Priority Research Program of CAS (Grant No. XDB07000000).

\end{document}